\documentclass[12pt,twoside,english]{article}
\usepackage[T1]{fontenc}
\usepackage[latin9]{inputenc}
\usepackage{geometry}
\usepackage{color}
\usepackage{babel}
\usepackage{units}
\usepackage[unicode=true,
 bookmarks=true,bookmarksnumbered=true,bookmarksopen=true,bookmarksopenlevel=1,
 breaklinks=false,pdfborder={0 0 1},backref=false,colorlinks=true]
 {hyperref}
\hypersetup{pdftitle={The LyX Tutorial},
 pdfauthor={LyX Team},
 pdfsubject={LyX-documentation Tutorial},
 pdfkeywords={LyX, documentation},
 linkcolor=black, citecolor=black, urlcolor=blue, filecolor=blue,pdfpagelayout=OneColumn, pdfnewwindow=true, pdfstartview=XYZ, plainpages=false}
\usepackage{breakurl}

\makeatletter
\usepackage{ifpdf} 
\ifpdf 

 \IfFileExists{lmodern.sty}{\usepackage{lmodern}}{}

\fi 

\let\myTOC\tableofcontents
\renewcommand\tableofcontents{%
  \frontmatter
  \pdfbookmark[1]{\contentsname}{}
  \myTOC
  \mainmatter }

\def\LyX{\texorpdfstring{%
  L\kern-.1667em\lower.25em\hbox{Y}\kern-.125emX\@}
  {LyX}}

\makeatother

\begin{document}

\title{Quantum Entanglement \\on Cosmological Scale}

\author{{\Large A. Nicolaidis and V. Kiosses}\medskip \\Theoretical Physics
Department\\Aristotle University of Thessaloniki\\54124 Thessaloniki,
Greece\\nicolaid@auth.gr}
\maketitle
\begin{abstract}
It has been suggested that relational logic, a form of logic developed
by C. S. Peirce, is the common inner syntax of quantum mechanics and
string theory. A relation may be represented by a spinor and the Cartan-Penrose
connection of spinor to geometry, allows to abstract geometry from
a calculus of relations-spinors, reviving Wheeler\textquoteright{}s
pregeometry. With a single spinor related to the null cone of Minkowski
space-time, we search for the geometry emerging when we entangle a
left-handed spinor and a right-handed spinor. We find that the quantum
entanglement generates an extra dimension and the distance in the
extra dimension is measured by the amount of entanglement. The emerging
geometry corresponds to a Milne space-time, with two branes coexisting
in the extra dimension. One brane hosts left-handed particles (our
brane), while the other brane hosts right-handed particles. A distinct
phenomenology accompanies our proposal. During the brane collision
all points are causally connected, making less pressing the inflationary
scenario. The left-right symmetry is achieved with having two \textquotedblleft{}mirror\textquotedblright{}
branes and the neutrino appears as the ideal mediator between the
branes. We may revisit also the dark matter, dark energy issues, with
everything on the other brane and in the bulk appearing \textquotedblleft{}dark\textquotedblright{}
to us. If biohomochirality is due to parity violation, we anticipate
that the life-forms in the other brane demonstrate opposite chiralities.
Our scheme brings closer logic - quantum theory - string theory \textendash{}
geometry \textendash{} cosmology and space-time, rather than an abstract
and a apriori construction, appears as the outcome of a quantum logical
act. 
\end{abstract}
\newpage

\subsubsection*{Introduction}

\qquad{}Everything is under evolution. The entire universe appears
as an ever-changing entity, with distinct stages in evolution. Ordinary
matter and radiation, stars, galaxies, clusters of galaxies emerge
as parts of an unfolding cosmic evolution. Next to the cosmological
evolution, the biological evolution follows and finally the human
language and culture appears. Do we dispose the necessary tools to
understand the pivotal process of universal evolution? Can we detect
and study the different modes of evolution? What kind of theory will
bind very different processes together and will reproduce the observed
time scales? Should we revisit the fundamental notions of space and
time and seek for a dynamical emergence of geometry and time?

\qquad{}In another direction, there is the unification drive in physics,
the effort of putting together within a unified framework distinct
theoretical approaches. Along the unification of terrestial and celestial
Mechanics (Newton in the $17^{\mathtt{th}}$ century), of optics with
the theories of electricity and magnetism (Maxwell in the $19^{\mathtt{th}}$
century), of spacetime geometry and gravity (Einstein in 1916), the
next step in this process is the unification of quantum mechanics
and general relativity. String theory, in this respect, appears as
a promising example of unification. Not unrelated is the quest for
the foundations and the internal architecture of a theory. Different
theories may share the same internal architecture or syntax, thus
reaching a deeper unified comprehension.

\qquad{}It has been suggested \cite{key-1} that relational logic,
a form of logic developed by C. S. Peirce during the years 1870-1880
\cite{key-2,key-3}, may offer new insights into these difficult issues.
Algebraic logic attempts to express the laws of thought in the form
of mathematical equations, and Peirce introduced relation as the irreducible
primary datum. All other terms or objects are defined in terms of
relations, transformations, arrows, morphisms. A relation $R_{ij}$
may receive multiple interpretations: as the proof of the logical
proposition $i$ starting from the logical premise $j$, as a transition
from the $j$ state to the $i$ state, as a measurement process that
rejects all impinging systems except those in the state $j$ and permits
only systems in the state $i$ to emerge from the apparatus. At the
core of the Peircean logical system is the composition of relations.
Two relations of the form $R_{ij}$, $R_{jk}$ may be composed giving
rise to a third transitive relation $R_{ik}$. It has been indicated
that the logical structure developed by Peirce bears great resemblance
to category theory, a remarkably rich branch of mathematics developed
by Eilenberg and Maclane in 1945 \cite{key-4}. A consistent effort
of categorification of physics has been developed already {[}5-7{]}. 

\qquad{}The symmetries and the dynamics of the Peircean relations
lead to the essential laws of quantum mechanics (the probability rule,
the commutation rules), indicating relational logic as the conceptual
foundation of quantum mechanics \cite{key-1}. In another direction,
when we adopt a double line representation for relations, the composition
rule for relations appears as string joining or string splitting \cite{key-1}.
Repetition of the {}``cubic-string'' interaction leads to geometrical
patterns reminiscent of Regge's discrete version of Riemannian geometry
\cite{key-8}, of Penrose's spin networks \cite{key-9}, of simplicial
quantum gravity \cite{key-10}, of string foam models \cite{key-11}.
A relation standing for a logical proof, may be represented by a spinor.
Consider for example the simplified case of two propositions $\left(i,j=1,2\right)$.
The emerging algebra is $SU\left(2\right)$, thus the underlying dynamics
is similar to a {}``spin 1/2 particle''. A proposed logical proof
(proposition $j$ entails proposition $i$) receives as an answer
a {}``yes'' or a {}``no'' statement, resembling a spinor, which
under measurement is revealed as {}``spin up'' or {}``spin down''
\cite{key-1,key-12}. We understand that a spin network may represent
the proof of an entire theorem in logic. Our approach goes in parallel
and vindicates Wheeler's vision. Wheeler pointed out that the genuine
explanations about the nature of something do not come about by explicating
a concept in terms of similar ones, but by reducing it to a different,
more basic kind of object. In this spirit Wheeler considered that
geometry is preceded by pregeometry, which is based on the calculus
of propositions \cite{key-13,key-14}. We suggest that this calculus
of propositions is governed by the rules of quantum relational logic.
It should be added that relations have a dynamical character and relational
logic may provide the appropriate framework to study and analyze evolutionary
dynamics. 

\qquad{}The profound connection between spinors and geometry was
established a hundred years ago by Cartan, who introduced spinors
as representation of the rotation group \cite{key-15}. Penrose used
spinor as the building block of discrete space-time and as a powerful
tool to study physics issues \cite{key-16}. Along this line we have
used spinors to represent a string worldsheet or an $AdS_{3}$ space
\cite{key-12}. In the present work we further extend the Cartan-Penrose
argument. If a single spinor, represented by a point on a Riemann
sphere, is connected to the Minkowski null cone, we may search for
the geometry emerging when two propositions-spinors are combined. 

\qquad{}A spinor may be represented by a Majorana field or a Dirac
field. We find that the composition of two left-handed fields, following
Majorana\textquoteright{}s instructions, leads to an entangled state
respecting the well known entanglement rules. On the other hand the
composition of a left-handed spinor and a right-handed spinor, in
Dirac\textquoteright{}s way, generates an extra dimension and the
emerging geometry corresponds to a Milne universe. In the third section
we examine the phenomenology of the proposed Milne universe, a universe
made out of two branes coexisting in the extra dimension.We point
out the similarity with the ekpyrotic scenario {[}17-19{]} and the
distinct difference, since within our model a brane (ours) hosts left-handed
fields, while the other brane hosts right-handed fields. We address
the issue of uniform behavior across the brane and we find that the
need for an inflationary scenario, to secure homogeneity and isotropy,
is less pressing. We study also the symmetries of the Milne universe
(especially the left-right symmetry), the mediation between the two
branes and we reconsider the subject of dark matter, dark energy within
the new framework. Tentatively we examine the handedness inherent
in our model and the homochirality observed in biology. 

\qquad{}Overall we suggest a new scheme which brings together logic,
quantum mechanics, string theory, geometry. Space-time is created
out of a quantum entanglement process and quantum logic governs not
only subatomic physics but cosmos itself. In the conclusions we summarize
our findings and indicate directions for future research.

\subsection*{Entanglement and geometry}

\qquad{}Consider the proposition-spinor $\left|u\right\rangle =\left(\begin{array}{c}
\xi\\
\eta
\end{array}\right)$. The relationship $R$ is defined by 
\begin{equation}
R=\left|u\right\rangle \left\langle u\right|=\left(\begin{array}{cc}
\xi\xi^{*} & \xi\eta^{*}\\
\eta\xi^{*} & \eta\eta^{*}
\end{array}\right)
\end{equation}
$R$ receives the decomposition 
\begin{equation}
R=\frac{1}{2}\sum_{\mu=0}^{3}X_{\mu}\sigma^{\mu}
\end{equation}
with $\sigma^{0}=\mathbf{1}$ and $\sigma^{i}$ $\left(i=1,2,3\right)$
the Pauli matrices. We deduce that 
\begin{equation}
X_{\mu}=\left\langle u\right|\sigma_{\mu}\left|u\right\rangle .
\end{equation}
$X_{\mu}$ satisfies identically the equation 
\begin{equation}
X_{1}^{2}+X_{2}^{2}+X_{3}^{2}-X_{0}^{2}=0\label{eq:4}
\end{equation}
Eq.(\ref{eq:4}) receives a double interpretation. It may represent
a null vector belonging to Minkowski spacetime \cite{key-15,key-16},
indicating the logic-algebraic origin of Minkowski spacetime. On the
other hand, with $X_{0}=1$, it represents the spinor Riemann-Bloch
sphere. Inversely, given eq.(\ref{eq:4}), we may search for its spinorial
representations. We obtain the Cartan-Weyl equations
\begin{eqnarray}
\left(\vec{X}\cdot\sigma-X_{0}\right)\left|u_{L}\right\rangle  & = & 0\\
\left(\vec{X}\cdot\sigma+X_{0}\right)\left|u_{R}\right\rangle  & = & 0
\end{eqnarray}
with $\left|u_{L}\right\rangle $ and $\left|u_{R}\right\rangle $
the left-handed and right-handed Weyl spinors. 

\qquad{}With a single spinor related to Minkowski spacetime, we may
ask what kind of geometry emerges when we entangle two spinors. Lets
consider first the Majorana type entanglement. Given a left-handed
spinor $\left|\psi_{L}\right\rangle $, following Majorana's recipe
\cite{key-20,key-21}, we may construct a right-handed spinor $\left|\chi_{R}\right\rangle $
by 
\begin{equation}
\left|\chi_{R}\right\rangle =\sigma_{2}\left|\psi_{L}\right\rangle ^{*}
\end{equation}
Starting with two left-handed Weyl spinors 
\begin{equation}
\left|\chi_{L}\right\rangle =\left(\begin{array}{c}
a\\
b
\end{array}\right)\qquad\left|\psi_{L}\right\rangle =\left(\begin{array}{c}
c\\
d
\end{array}\right)
\end{equation}
we define the four-component Majorana spinor 
\begin{equation}
\left|\Psi_{M}\right\rangle =\left(\begin{array}{c}
\left|\chi_{L}\right\rangle \\
\sigma_{2}\left|\psi_{L}\right\rangle ^{*}
\end{array}\right).
\end{equation}
With $\left\langle \Psi_{M}\right|=\left|\Psi_{M}\right\rangle ^{\dagger}\gamma_{0}$
we find
\begin{eqnarray}
X_{1}=\left\langle \Psi_{M}\right.\left|\gamma_{1}\right|\left.\Psi_{M}\right\rangle  & = & a^{*}b+b^{*}a-cd^{*}-dc^{*}\\
X_{2}=\left\langle \Psi_{M}\right.\left|\gamma_{2}\right|\left.\Psi_{M}\right\rangle  & = & i\left(-a^{*}b+b^{*}a-cd^{*}+dc^{*}\right)\\
X_{3}=\left\langle \Psi_{M}\right.\left|\gamma_{3}\right|\left.\Psi_{M}\right\rangle  & = & a^{*}a-b^{*}b-cc^{*}+dd^{*}\\
X_{0}=\left\langle \Psi_{M}\right.\left|\gamma_{0}\right|\left.\Psi_{M}\right\rangle  & = & -a^{*}a-b^{*}b+cc^{*}+dd^{*}
\end{eqnarray}
where we used
\begin{equation}
\gamma_{\mu}:\,\gamma_{i}=\left(\begin{array}{cc}
0 & \sigma^{i}\\
\sigma^{i} & 0
\end{array}\right),\;\gamma_{0}=\left(\begin{array}{cc}
0 & \mathbf{1}_{2}\\
-\mathbf{1}_{2} & 0
\end{array}\right).
\end{equation}
The quantity $X_{1}^{2}+X_{2}^{2}+X_{3}^{2}-X_{0}^{2}$ is not anymore
zero. We find
\begin{equation}
X_{1}^{2}+X_{2}^{2}+X_{3}^{2}-X_{0}^{2}\equiv M_{M}^{2}=4\left|\left(ad-cb\right)\right|^{2}
\end{equation}
Considering
\begin{eqnarray*}
X_{4}=i\left\langle \Psi_{M}\right|\left.\Psi_{M}\right\rangle  & = & -2i\, Im\left(ad-cb\right)\\
X_{5}=\left\langle \Psi_{M}\right.\left|\gamma_{5}\right|\left.\Psi_{M}\right\rangle  & = & 2i\, Re\left(ad-cb\right)
\end{eqnarray*}
with
\begin{equation}
\gamma_{5}=-i\gamma_{0}\gamma_{1}\gamma_{2}\gamma_{3}=\left(\begin{array}{cc}
\mathbf{1}_{2} & 0\\
0 & -\mathbf{1}_{2}
\end{array}\right),
\end{equation}
we obtain
\begin{equation}
M_{M}^{2}=-\left(X_{4}^{2}+X_{5}^{2}\right).
\end{equation}
The requirement that the Riemann-Bloch sphere remains intact, equ.(\ref{eq:4}),
implies the condition 
\begin{equation}
ad-cb=0\label{eq:18}
\end{equation}
On the other hand a generic two-qubit state is written as 
\begin{equation}
\left|\Psi\right\rangle =a\left|00\right\rangle +b\left|01\right\rangle +c\left|10\right\rangle +d\left|11\right\rangle 
\end{equation}
The state $\left|\Psi\right\rangle $ is {}``separable'', i.e. it
can be written as a simple product of individual kets, when the same
condition (\ref{eq:18}) is satisfied. Thus our Majorana entanglement
reproduces the ordinary two-qubit entanglement, usually obtained by
making appeal to the quaternion formalism \cite{key-22,key-23}.

\qquad{}For the Dirac entanglement we select a left-handed Weyl spinor
and a right-handed Weyl spinor. Writing 
\begin{equation}
\left|\chi_{L}\right\rangle =\left(\begin{array}{c}
a\\
b
\end{array}\right)\qquad\left|\psi_{R}\right\rangle =\left(\begin{array}{c}
c\\
d
\end{array}\right)
\end{equation}
and
\begin{equation}
\left|\Psi_{D}\right\rangle =\left(\begin{array}{c}
\left|\chi_{L}\right\rangle \\
\left|\psi_{R}\right\rangle 
\end{array}\right)
\end{equation}
we obtain 
\begin{eqnarray}
X_{1}=\left\langle \Psi_{D}\right.\left|\gamma_{1}\right|\left.\Psi_{D}\right\rangle  & = & a^{*}b+b^{*}a-c^{*}d-d^{*}c\\
X_{2}=\left\langle \Psi_{D}\right.\left|\gamma_{2}\right|\left.\Psi_{D}\right\rangle  & = & i\left(-a^{*}b+b^{*}a+c^{*}d-d^{*}c\right)\\
X_{3}=\left\langle \Psi_{D}\right.\left|\gamma_{3}\right|\left.\Psi_{D}\right\rangle  & = & \left|a\right|^{2}-\left|b\right|^{2}-\left|c\right|^{2}+\left|d\right|^{2}\\
X_{0}=\left\langle \Psi_{D}\right.\left|\gamma_{0}\right|\left.\Psi_{D}\right\rangle  & = & -\left(\left|a\right|^{2}+\left|b\right|^{2}\right)-\left(\left|c\right|^{2}+\left|d\right|^{2}\right)
\end{eqnarray}
The quantity $X_{1}^{2}+X_{2}^{2}+X_{3}^{2}-X_{0}^{2}$ is not anymore
zero. We obtain 
\begin{equation}
X_{1}^{2}+X_{2}^{2}+X_{3}^{2}-X_{0}^{2}\equiv-M_{D}^{2}=-4\left|\left(a^{*}c+b^{*}d\right)\right|^{2}\label{eq:26}
\end{equation}
Considering 
\begin{eqnarray*}
X_{4}=i\left\langle \Psi_{D}\right|\left.\Psi_{D}\right\rangle  & = & -2\, Im\left(a^{*}c+b^{*}d\right)\\
X_{5}=\left\langle \Psi_{D}\right.\left|\gamma_{5}\right|\left.\Psi_{D}\right\rangle  & = & -2\, Re\left(a^{*}c+b^{*}d\right)
\end{eqnarray*}
we find
\begin{equation}
M_{D}^{2}=\left(X_{4}^{2}+X_{5}^{2}\right)
\end{equation}
The absence of entanglement and the re-establishement of the Riemann-Bloch
sphere is obtained, when the condition 
\begin{equation}
a^{*}c+b^{*}d=0
\end{equation}
is fulfilled.

\qquad{}Let us define $T=X_{0}$, $t=M_{D}$. The Dirac entanglement,
equ.(\ref{eq:26}), takes then the form of a space-like hyperboloid
\begin{equation}
T^{2}-\sum_{i=1}^{3}X_{i}^{2}=t^{2}
\end{equation}
The above constraint is automatically satisfied by 
\begin{equation}
T=t\,\cosh\rho\quad X_{i}=n_{i}\, t\,\sinh\rho\quad\sum_{i=1}^{3}n_{i}^{2}=1
\end{equation}
The amount of quantum entanglement is determined by $t$. The topological
equivalence between the Riemann-Bloch sphere and the null cone of
Minkowski space is obtained by letting the radius of the Riemann-Bloch
sphere becoming the continuous time $X_{0}$. In a similar fashion
we consider $t$ as a continuous variable. The induced metric becomes
\begin{eqnarray*}
ds^{2} & = & dT^{2}-\sum_{i=1}^{3}dX_{i}^{2}\\
 & = & dt^{2}-t^{2}\left(d\rho\right)^{2}-t^{2}\sinh^{2}\rho\, d\Omega_{2}^{2}
\end{eqnarray*}
This represents a Milne universe. Within our approach, geometry and
space-time, rather than abstract and a priori mathematical constructions,
emerge as the outcome of a quantum act, the act of quantum entanglement.
In the next section we will examine the phenomenological implications
of a Milne universe.

\subsection*{Phenomenology of a Milne Universe}

\qquad{}Our starting point is logic, relational logic as a foundation
of quantum mechanics and string theory. The composition rule of relations
builds up both geometry and quantum mechanics. This development gives
further credit to the notion that a quantum system may encode a logical
proof \cite{key-25}. Space-time itself is sought as an emergent property
of a deeper theory or, to use Wheeler\textquoteright{}s terminology,
a pregeometry \cite{key-14}. The close connection we found between
a relation and a spinor, led us to follow the Cartan-Penrose argumentation
\cite{key-9,key-15,key-16} and use the relation-spinor as the building
block of space-time. If a single spinor is associated to Minkowski
space-time, our study indicates that the entanglement of two spinors
gives rise to a more complex space-time. At first quantum entanglement
generates an extra dimension, indicating that a quantum system in
$d$ dimensions is analogous to a classical system in $d+1$ dimensions
\cite{key-26}. Further the emerging metric corresponds to a Milne
space-time, with two $3$-dimensional branes coexisting in the extra
dimension. The outcome is very similar to the ekpyrotic model {[}17-19{]}.
According to the ekpyrotic scenario a violent collision between the
two branes results to a conflagration, resembling the conventional
\textquotedblleft{}big bang\textquotedblright{}. It should be noted
though that the ekpyrotic model is derived within the heterotic M-theory
\cite{key-27}, while the Milne universe we suggest emerges out of
a quantum logical process. Furthermore the two branes are not identical
in our case. By construction one brane hosts left-handed particles
(our brane), while the other brane hosts right-handed particles.

\qquad{}The Lagrangian for a particle in a Milne universe is
\begin{equation}
L=\frac{1}{2}\left[\dot{t}^{2}-t^{2}\dot{\varrho}^{2}-t^{2}\sinh^{2}\varrho\left(\dot{n}_{1}^{2}+\dot{n}_{2}^{2}+\dot{n}_{3}^{2}\right)\right]
\end{equation}
The equations of motion are easily solved in the case of one-dimensional
branes. We find
\begin{equation}
\varrho\simeq-\log t
\end{equation}
As $t$ goes to zero we approach the time of brane-collision (the
ekpyrosis moment) and the distance $\varrho$ across the brane becomes
infinite. We gather that at the collision time the correlation along
the brane is infinite and all the points in the brane are causally
connected \cite{key-28}. The apparent homogeneity and isotropy of
our universe can be accounted therefore, making less pressing the
need for an inflantionary scenario \cite{key-29}.

\qquad{}The conventional way to restore left-right symmetry is to
introduce an extra $SU\left(2\right)_{R}$ gauge group in the energy
desert above the scale of the standard $SU\left(2\right)_{L}$ interactions.
The right-handed gauge bosons are more massive compared to the left-handed
gauge bosons, leading to parity violation at low energies \cite{key-30,key-31}.
Within our approach the left-right symmetry is achieved with the extra
dimension hosting two {}``mirror'' branes, a left-handed brane and
a right-handed brane. Higgs scalars, denoted by $\phi_{L}$ and $\phi_{R}$,
live in the corresponding branes, though having different vacuum expectation
values. An interaction term $\lambda\phi_{L}^{2}\phi_{R}^{2}$ may
induce a mixing of the two Higgs scalars, serving also as a mediator
between the two branes. At the ekpyrotic moment $\left(t=0\right)$,
the full conformal $\left(L+R\right)$ symmetry is achieved. From
a phenomenological point of view the particles living in the {}``other''
brane behave as mirror-duplicate of the particles in our visible world
\cite{key-32,key-33}.

\qquad{}The most prominent candidate for mediation between the two
branes is the neutrino particle. The left-handed neutrino, an essential
ingredient of the standard model, resides in our brane, while its
counterpart, the right-handed neutrino, resides in the other brane.
The two braneworlds are equivalent to a two-sheeted spacetime $M_{4}\times Z_{2}$
\cite{key-33,key-34}, with $M_{4}$ standing for a four-dimensional
continuous manifold and the fifth dimension reduced to two discrete
points separated by a distance $2t$ (the amount of entanglement).
In a five dimensional $Z_{2}$-Dirac equation, the neutrino mass will
appear as a term connecting the two branes and neutrino oscillations
will acquire the novel form of swapping between the branes. Our scheme
offers a natural explanation for the neutrino masses. The neutrino
mass is at the same time the entanglement between two spinors and
the distance between the two branes. The tiny neutrino masses reflect
the small distance separating the branes. Notice also the evolving
nature of the neutrino mass. Close to ekpyrosis the Dirac neutrino
is massless, affecting in a distinc way cosmology.

\qquad{}In our universe we register three main components. Visible
matter accounts for about $4\%$, dark matter for $21\%$ and dark
energy for the remainder. In the usual approach we consider that dark
matter consists of particles (classified as cold, warm, hot), while
the dynamics of dark energy is assumed by the cosmological constant
$\Lambda$. In our case we explain dark matter and dark energy by
invoking geometry. Everything that is localized in the other brane
and in the extra dimension appears dark to us. If we assume that in
our observed brane, matter is equally shared between visible and invisible
matter, with the same analogy holding in the other brane, then we
obtain naturally that our visible matter is $\sim\nicefrac{1}{5}$
of the entire matter content, in agreement with the observations.
Similar arguments have been developed in models where the brane is
folded many times inside the extra dimension \cite{key-35}. Since
we have located most of the matter outside our brane, we may easily
imagine that the local distribution of visible matter is not identical
to the local distribution of the entire matter. In an imaginative
journey, as we leave the center of our galaxy, moving a fraction of
the millimeter across the extra dimension to reach the {}``other''
brane, we are not going to land at the center of the {}``other''
galaxy. The local matter distribution in the {}``mirror'' brane
is not the same with the matter distribution in our brane, and the
net effect is a displacement of the dark matter density with respect
the visible matter density. Such an effect has been recently observed.
The peak of the local dark matter density differs from the center
of our galaxy by several hundred parsec \cite{key-36}.

\qquad{}Gravity is not confined to the brane and matter fields on
the brane can emit gravitational waves into the bulk. The brane energy-momentum
tensor is not conserved therefore and it is essential to include this
energy flow into realistic cosmological models. The leakage of gravitational
energy into extra dimension affects the cosmological expansion \cite{key-37},
inducing an accelerating expansion for the three dimensional space
\cite{key-37-2}. Thus an alternative to the non-zero cosmological
constant scenario is offered.

\qquad{}It is well known that life manifests a strong biohomochirality.
In the living organisms we encounter only left-handed aminoacids and
right-handed sugars. There are many hypotheses advanced regarding
the origin of biohomochirality. We would like to point out the hypothesis
linking the bio-asymmetry to the fundamental parity violation of the
weak interactions. The weak neutral currents (mediated by $Z^{0}$)
stabilize preferentially the L-aminoacids and D-sugars over the D-aminoacids
and L-sugars. The difference in energy between the two enantiomers
is small. A phenomenon of autocatalysis in a far from equilibrium
state, amplifies the small difference and in a period of $10^{4}$
years leads to a unique chirality \cite{key-38}. If this working
hypothesis is a valid one, then in the {}``mirror'' brane the corresponding
weak neutral currents will operate in the opposite direction. In that
case the life-forms of the other brane will have exclusively D-aminoacids
and L-sugars.

\subsection*{In lieu of conclusions }

\qquad{}We are used first to wonder about particles or states and
then about their interactions. First to ask about \textquotedblleft{}what
is it\textquotedblright{} and afterwards \textquotedblleft{}how is
it\textquotedblright{}. On the other hand, quantum mechanics and string
theory display a highly relational nature. We are led to reorient
our thinking and consider that things have no meaning in themselves,
and that only the correlations between them are \textquotedblleft{}real\textquotedblright{}
\cite{key-39}. We adopted the Peircean relational logic as a consistent
framework to prime correlations and gain new insights into these theories.
Our representation of a relation by a spinor allowed us to connect
logic with geometry and space-time. We found that the entanglement
of two spinors generates an extra dimension and the distance in the
extra dimension is measured by the amount of entanglement. Finally
two brane-geometries are entangled together: one brane hosting left-handed
particles and another brane hosting right-handed particles. We would
like to draw also a parallel with the CFT/AdS duality \cite{key-40,key-41}.
A spinor may be viewed as the space-time orientation of a pixel on
a holographic screen \cite{key-42}. Through the pixel on the light-cone
a null ray passes. What the quantum entanglement offers us is the
exploration of the internal space of the light-cone (the bulk). On
very general grounds it has been shown the correspondence between
quantum entanglement, classical renormalization group and holographic
gauge/gravity duality {[}43-46{]}. The stringy degrees of freedom
operating in the CFT/AdS duality are assumed in our case by the entanglement
degrees of freedom and quantum phenomena are encoded in classical
geometry.

\qquad{}The emerging geometry, a Milne space-time with two chirally
oriented branes, offers a rich phenomenology: a novel way to approach
left-right symmetry on cosmological scale, a link of dark matter and
dark energy to the dynamics of extra space, particles and forces as
mediators between the branes, a connection between the left-right
symmetry breaking and biohomochirality. Within our approach time acquires
a new role. The beginning of time is not set at the \textquotedblleft{}big
bang\textquotedblright{}, 13.7 billions years ago, but somehow is
lost in the previous aeons. The time allocated to universe\textquoteright{}s
evolution is longer than 13.7 billion years, involving also the time
periods between successive \textquotedblleft{}ekpyrotic\textquotedblright{}
moments. During these periods-aeons part of the creation of the universe
was carried out. Thus the difficulty in accommodating the different
distinct phenomena within a single history, exemplified by the account
of the \textquotedblleft{}anthropic principle\textquotedblright{}
\cite{key-48}, is significantly reduced. Searching for the ruins
from a previous aeon, or the archeology of the universe, is the most
intricate and complicated task. Yet, it has been already suggested
that in the CMB there are traces from an activity preceding the \textquotedblleft{}big-bang\textquotedblright{}
\cite{key-49}.

\qquad{}Altogether the new syntax we have introduced brings closer
logic \textendash{} quantum mechanics \textendash{} string theory
- cosmology and allows addressing foundational questions regarding
the evolution of the universe. A relevant phenomenology may serve
as the testing ground of these ideas.

\section*{Acknowledgments}

One of us (A.N.) would like to acknowledge useful discussions with
Prof. Brian Schmidt and Prof. Ernest Ma during the $16^{th}$ Paris
Cosmology Colloquium. The present work is supported by the Templeton
Foundation.

\vfill{}
\vfill{}

\end{document}